\documentclass[fleqn,usenatbib]{mnras}

\usepackage[T1]{fontenc}

\DeclareRobustCommand{\VAN}[3]{#2}
\let\VANthebibliography\thebibliography
\def\thebibliography{\DeclareRobustCommand{\VAN}[3]{##3}\VANthebibliography}

\usepackage{graphicx}	
\usepackage{amsmath}	
\usepackage{amssymb}	

\usepackage{pdfpages}
\usepackage{adjustbox}
\usepackage{multirow}
\usepackage[usestackEOL]{stackengine}
\usepackage{ulem}

\usepackage{newtxtext,newtxmath}

\newcommand{\kms}{\,km\,s$^{-1}$}
\newcommand{\ms}{M$_{\sun}$}

\newcommand{\aFe}{\mbox{$\mbox{[$\alpha$/Fe]}$}}

\definecolor{m-purple}{HTML}{7e1e9c}
\definecolor{m-green}{rgb}{0.0, 0.5, 0.0}
\definecolor{m-orange}{HTML}{ff7f00}

\defcitealias{eftekhari2021}{E21}
\defcitealias{eftekhari2022}{E22}
\defcitealias{labarbera2019}{LB19}
\defcitealias{martin2015d}{MN15}
\defcitealias{conroy2018}{C18}

\title[CO Absorptions in NGC~1277]{The relic galaxy NGC~1277 rules out intermediate-age stellar populations origin of CO-strong absorptions in massive early-type galaxies}

\author[E. Eftekhari et al.]{
Elham Eftekhari,$^{1,2,3}$\thanks{E-mail: elhamea@ipm.ir }
Francesco La Barbera,$^{4}$
Alexandre Vazdekis,$^{2,3}$
and Michael Beasley$^{2,3}$
\\
$^{1}$School of Astronomy, Institute for Research in Fundamental Sciences - IPM, 19395-5531, Tehran, Iran\\
$^{2}$Instituto de Astrof{\'{\i}}sica de Canarias, E-38200 La Laguna, Tenerife, Spain\\
$^{3}$Departamento de Astrof{\'{\i}}sica, Universidad de La Laguna, E-38205 La Laguna, Tenerife, Spain\\
$^{4}$INAF-Osservatorio Astronomico di Capodimonte, sal. Moiariello 16, Napoli I-80131, Italy\\
}

\date{Accepted 2022 June 22. Received 2022 June 19; in original form 2022 April 12}

\pubyear{2022}

\begin{document}
\label{firstpage}
\pagerange{\pageref{firstpage}--\pageref{lastpage}}
\maketitle

\begin{abstract}

Massive Early-Type Galaxies (ETGs) show several strong CO absorption features in their \textit{H}- and \textit{K}-band spectra that cannot be explained by state-of-the-art stellar population models. For many years, the disagreement has been attributed to the presence of intermediate-age stellar components that are dominated by stars in the Asymptotic Giant Branch (AGB) phase. However, no robust evidence of this scenario has been provided so far. One way to test this claim is by comparison of CO indices for ETGs and for relic galaxies. Lacking the intermediate-age stellar populations, relic galaxies provide us with a unique opportunity to address the origin of strong CO absorptions in ETGs. Here, we utilize the prototype relic galaxy NGC~1277 and compare the CO absorption features of this galaxy with the ones of a representative sample of massive ETGs. We show that the CO lines in both systems have similar strengths, significantly stronger than the predictions of stellar population synthesis models. We conclude that intermediate-age stellar populations in massive ETGs are not the culprit of the strong CO absorptions.

\end{abstract}

\begin{keywords}
infrared: galaxies  -- galaxies: stellar content 
\end{keywords}



\section{Introduction} \label{sec:sec1}

A puzzling observational result has been emerged from the observation of CO bandhead in the \textit{K} band since 1990s \citep[e.g.][]{mobasher1996,james1999}, suggesting that the strong CO absorption in Early-Type Galaxies (ETGs) might be due to the presence of Asymptotic Giant Branch (AGB) stars, from intermediate-age populations. This would imply that most massive ETGs do not evolve completely passively after the bulk of their star formation has stopped, but may have experienced new star formation episodes with extended Star Formation Histories (SFHs). This is in sharp contrast with results obtained in the optical spectral range. In this Letter, we take advantage of the deep spectroscopic observations of the CO overtone bands in both \textit{H} and \textit{K} spectral windows to test this scenario with an empirical approach. To this aim, we compare the strengths of CO indices, defined in \citet{eftekhari2021} (hereafter \citetalias{eftekhari2021}), in the ``genuine" relic galaxy NGC~1277~\citep{trujillo2014,ferre2015,martin2015d,ferre2017}, with the stellar population models.  A recent study on CO absorption features in massive ETGs shows that stellar population synthesis (SPS) models predict systematically weaker CO absorptions than that observed in massive ETGs \citep{eftekhari2022} (hereafter \citetalias{eftekhari2022}). As relic galaxies are massive galaxies that did not experience a significant merger event since their early collapse phase, they are perfect laboratories to study the in-situ stellar component of massive ETGs. Old ages of the bulk of the stars in NGC~1277 is confirmed by studies in both optical \citep{trujillo2014,ferre2015,martin2015d} and Near-UltraViolet (NUV) (Salvador-Rusi{\~n}ol et al. submitted) windows. Therefore, any mismatch between the observations of CO bandhead in this galaxy with the stellar population models can not be due to the intermediate/young populations (e.g. Thermaly Pulsating(TP)-AGB) that dominate the light in the Near-InfraRed (NIR), discarding the attribution of strong CO absorptions in massive ETGs to an emphasised contribution from AGB stars.

This Letter is structured as follows: We describe our samples and data in Sec.~\ref{sec:sec2}, and the stellar population  models used in this Letter in Sec.~\ref{sec:sec3}. We compare the spectra of our samples with models' predictions around CO absorptions in Sec.~\ref{sec:sec4.1}, while a quantitative comparison of CO line-strengths is given in Sec.~\ref{sec:sec4.2}. In Sec.~\ref{sec:sec5}, we discuss the results.

\section{Samples and Data}\label{sec:sec2}

We use a sample of seven massive ETGs from \citet{labarbera2019} (hereafter \citetalias{labarbera2019}) and compare it with NGC~1277, which is regarded as a prototype relic galaxy. Note that all galaxies in these samples have a similar stellar mass of $\sim$10$^{11}$\ms. In the following, we provide a summary of the main properties of our samples and data.

    \subsection{Massive ETGs}\label{sec:sec2.1}
    
    The sample of extremely massive ($\sigma$>300~\kms) ETGs from \citetalias{labarbera2019} comprises seven galaxies at redshift around 0.05 drawn from the SPIDER Survey \citep{labarbera2010} and SDSS DR7, intended to be representative of the high-mass end population of ETGs. \citetalias{labarbera2019} obtained long-slit spectra with high signal-to-noise ratio (SNR) (>170~\AA$^{-1}$) and $\sim$5500 spectral resolution, with the VLT/X-SHOOTER spectrograph \citep{vernet2011}.
    
    The high-quality spectra of galaxies from \citetalias{labarbera2019} (hereafter XSGs) have been extracted within two apertures: one matching the size of the aperture for NGC~1277, i.e. 1.2~kpc (see below), and the other one within 1R$_{\rm e}$. Note that we used the average effective radius (R$_{\rm e}$$\sim$3.9~kpc) of the bulge component of these galaxies obtained with the most sensible decomposition method for XSGs (see appendix C of \citetalias{labarbera2019}). Also note that in this Letter we rely on the stacked spectrum of XSGs although we show the scatter in CO strengths from the individual XSGs.
    
    \subsection{The relic galaxy}\label{sec:sec2.2}
    
    We compare the CO indices of our sample of giant ETGs with those for the archetypal relic galaxy, NGC~1277.  NGC~1277 is a massive (M$_{\star}=1.2\pm0.4\times10^{11}$\ms), compact (R$_{\rm e}=1.2$~kpc) and old ($>$~10~Gyr) system, with a central metallicity of [Fe/H]~$=0.20\pm0.04$ and $\alpha$ abundance of \aFe\ $=+0.4\pm0.1$ \citep{trujillo2014}. It shows a bottom-heavy Initial Mass Function (IMF) with a slope of $\Gamma_{b}$$\sim$3 in the center and a very mild IMF radial gradient (\citet{martin2015d}; hereafter \citetalias{martin2015d}).

    \textit{H}- and \textit{K}-band data for NGC~1277 were obtained with EMIR spectrograph at the Gran Telescopio Canarias (a 10.4~m telescope at the Observatorio del Roque de Los Muchachos, La Palma) during 6 different nights in 2017 and 2019 (Proposal ID: GTC3-17B), with a 0.6\arcsec\ wide long-slit. Light was dispersed with the \textit{H} and \textit{K} grisms, with wavelength coverages from 1.5~$\mu$m to 1.8~$\mu$m and 2.0~$\mu$m to 2.4~$\mu$m with resolutions of $\sim$4300 and $\sim$4100, respectively. The spectra were recorded with galaxy offsets in an ABBA nodding pattern along the slit, where A and B are different positions on the slit, separated by 90\arcsec. A 120~s exposure was recorded at each slit position, and multiple ABBA cycles were repeated (9 and 12 cycles in \textit{H} and \textit{K} bands, respectively). The ABBA exposures used to remove sky emission lines by subtracting the pairs of images taken at different slit positions (2A - 2B). Note that we also removed any residual from sky emission lines, left over from the pair subtraction, by linear interpolation. Spectra of stars with different spectral type were obtained after each block of observations to monitor telluric absorption lines.  In the following, we summarize the main challenges that we encountered during data reduction.
    
    The EMIR dome flats turned to degrede the quality of the frames rather than improving it. Therefore, flat-fielding was not applied. Note that this is not going to affect our results, as we are only interested in a small region of the frames around the photometric center of the galaxy, where we do not expect significant long-scale flat-field variations. Moreover, the pixel-to-pixel variations are taken into account by the variance map, obtained by combining all dithered spectra.
    We applied corrections for telluric absorption lines by using the software Molecfit \citep{smette2015,kausch2015}, that constructs a synthetic telluric absorption model from the observed spectrum itself. Since the spectral resolution of EMIR slightly changes with  wavelength, we ran molecfit on different windows and then combined the corresponding telluric absorption models. Notice that we re-observed NGC~1277 on 2019, in order to significantly improve the flux calibration in the region of the K-band CO (see Sec.~\ref{sec:sec4.1}). 
    
\begin{figure*}
\centering
\includegraphics[width=\linewidth]{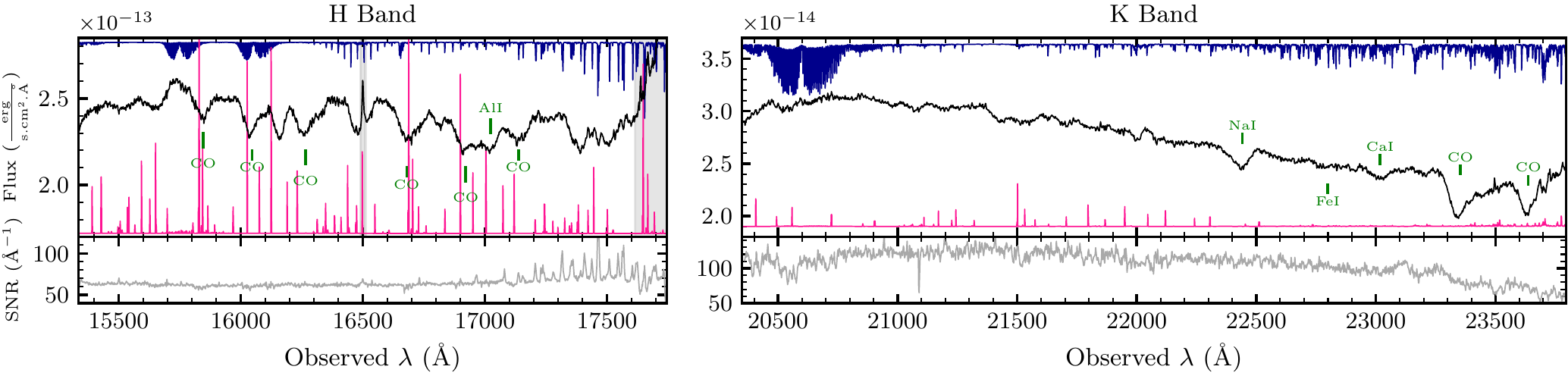}
\protect\caption[]{Reduced spectra of NGC~1277 in \textit{H} (left) and \textit{K} (right) bands. For each panel, the upper plot shows the NGC~1277 spectrum (black) extracted within an aperture of 1R$_{\rm e}$ and the rescaled telluric absorption (dark blue) and sky emission spectra (pink) obtained by the Skycalc tool \citep{noll2012,jones2013}. Lower panels plot  SNR spectra. }
\label{fig:fig1}
\end{figure*}

    Figure~\ref{fig:fig1} shows the reduced spectra of NGC~1277, extracted within an aperture of 1R$_{\rm e}$ (1.2~kpc), featuring a SNR limit of $\sim$60~\AA$^{-1}$ in \textit{H} band and about $\sim$100~\AA$^{-1}$ in \textit{K} band.
    
\section{Stellar population models}\label{sec:sec3}

We compare observed CO index strengths with predictions of E-MILES stellar population synthesis (SPS) models \citep{rock2015, vazdekis2016, rock2016}. We also use \citet{conroy2018} (hereafter \citetalias{conroy2018}) models  which provide predictions for the effect of carbon enhancement on  CO indices. A combined model of an AGB-enhanced plus an old population is also used in this Letter to show the effect of  an emphasized  contribution  of  AGB stars  to  E-MILES models. Moreover, we also use the two empirically corrected E-MILES models with solar metallicity, Milky Way(MW)-like and bottom-heavy IMF from \citetalias{eftekhari2022}. These models emphasise the contribution of CO-strong cool giant stars with respect to other stars with similar effective temperature.  A full description of the models used in this Letter can be found in \citetalias{eftekhari2022}.

\section{CO spectral indices}\label{sec:sec4}

In the following, we first compare the spectra of E-MILES models around CO bands at 1.6~$\mu$m and 2.3~$\mu$m to the stacked spectra of XSGs (one extracted within 1.2~kpc and  another one within 1R$_{\rm e}$) and the spectrum of NGC~1277 (extracted within 1R$_{\rm e}$=1.2~kpc) (Sec.~\ref{sec:sec4.1}). Then we compare model and observed  line-strengths in Sec.~\ref{sec:sec4.2}.

    \subsection{CO indices: observed vs models spectra}\label{sec:sec4.1}
    
\begin{figure*}
\centering
\includegraphics[width=0.99\linewidth]{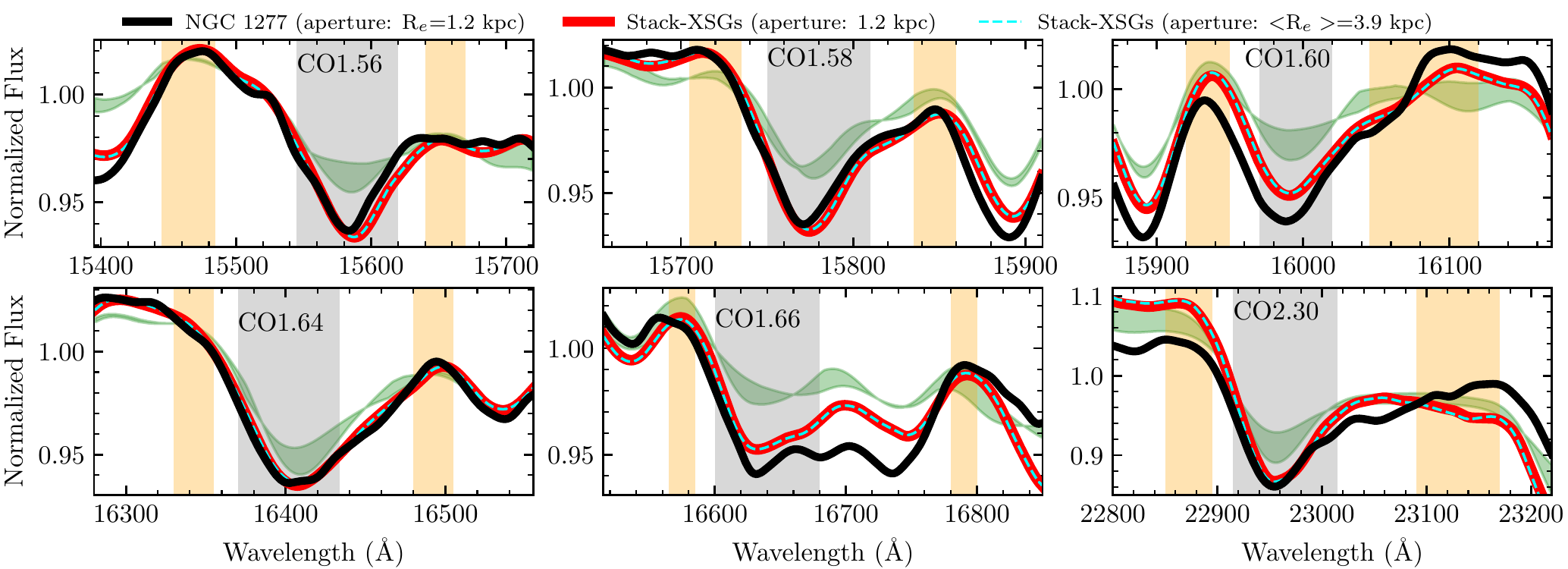}
\protect\caption[]{Spectral regions around the CO features, showing NGC~1277 (black), stacked XSGs (solid red and dashed cyan) and E-MILES models. The green region shows how the CO features change for E-MILES SSP models over a wide range of ages (from 1 to 14~Gyr) and metallicities (from -0.35 to +0.26~dex), and IMF slopes ($\Gamma_{b}$ from 0.3 to 3.5). The model and observed spectra have been convolved to a common resolution of $\sigma$=360~\kms. The central bandpasses of CO indices, as well as the blue and red pseudo-continua, are from \citetalias{eftekhari2021} and are shown as grey and orange areas, respectively. All spectra have been normalised to the mean flux within pseudo-continua bands. Remarkably, for all CO features, galaxies show stronger absorption than any model and the spectra of the relic galaxy and massive ETGs show similar depths in most cases.}
\label{fig:fig2}
\end{figure*}

    In Fig.~\ref{fig:fig2}, we show the stacked spectra of XSGs from \citetalias{labarbera2019} (solid red and dashed cyan lines) and the spectrum of NGC~1277 (black), around CO absorptions from \textit{H} throughout \textit{K} bands.  E-MILES model spectra are plotted as green shaded regions, showing the allowed range in CO strengths for a wide range of age (from 1 to 14~Gyr), metallicity (from -0.35 to +0.26~dex) and logarithmic bimodal IMF slope (from 0.3 to 3.5). The wavelength definitions of CO indices, from \citetalias{eftekhari2021}, are shown with shaded grey and orange areas, corresponding to indices' bandpass and pseudo-continua bands, respectively. The plots for CO1.60, CO1.66, and CO2.30 panels, suggest that there is likely an issue with  flux calibration in these regions, as the blue pseudo-continua in NGC~1277 are lower than the stack of XSGs and models, while the red pseuso-continua show the opposite behaviour. However, according to table 2 of \citetalias{eftekhari2021}, the impact of flux calibration on line-strength indices of CO1.60 and CO1.66 is negligible, while CO2.30  is among the indices with highest sensitivity to flux calibration. Actually, this is the reason why NGC~1277 was observed twice in K band on 2017 and 2019, with different stars being used for flux calibration. As a quantitative assessment, it is worth noting that the strength of the CO2.30 index changes around 1~\AA\ between the 2017 data and the final reduction (note that using jump-like, rather than Lick-style definition, (e.g.~\citet{marmol2008}) would make the measurements far more uncertain). In general, the disagreement between observed and model CO indices can be seen in all panels. The CO indices of the stacked spectra of XSGs and the NGC~1277 spectrum are much stronger than any Simple Stellar Population (SSP) model. Note that the stacked spectra of XSGs, with two different apertures, are almost identical, which indicates mild CO gradients. Moreover, the depths of CO indices for NGC~1277, except for CO1.60 and CO1.66, do not deviate significantly from those of the XSGs stacked spectra, suggesting that, overall,  there is no  difference in the stellar populations contributing to CO strengths of NGC~1277 and the general population of massive ETGs.
    
\subsection{CO line-strengths}\label{sec:sec4.2}
    
\begin{figure*}
\centering
\includegraphics[width=\linewidth]{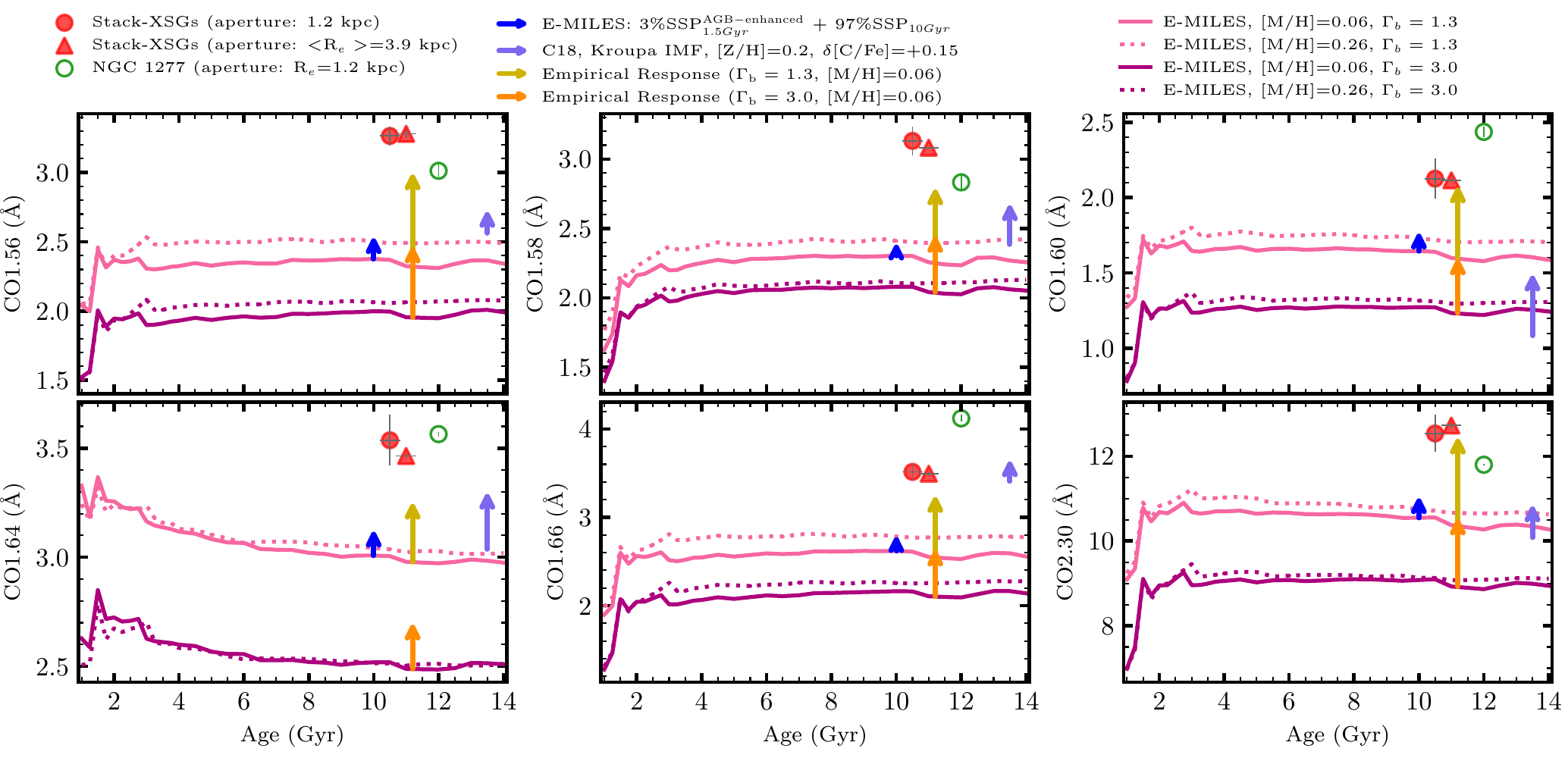}
\protect\caption[]{The CO line-strength indices in \textit{H} and \textit{K} bands as a function of age for models (lines) and observations (discrete points). These plots show that the observations have stronger CO absorption than predictions of the models. Red points are extremely massive ETGs and the green point is the massive relic galaxy, NGC~1277. As summarized in the top labels, the arrows show variations of the line-strengths, with respect to reference models, obtained by (i) adding  3\% of a young (1.5~Gyr) population  with emphasized contribution of AGB stars to an old (10~Gyr) SSP (blue arrow); (ii) enhancing [C/Fe] by +0.15~dex with \citetalias{conroy2018} models (violet arrow); (iii) considering models with an enhanced contribution of CO-strong giant stars, for a MW-like and bottom-heavy IMFs (yellow and khaki arrows, respectively). NGC~1277 and massive ETGs show similar line-strengths in most panels, significantly above the model predictions. Interemediate-age populations, not present in NGC~1277, cannot explain the observed CO mismatch.  
}
\label{fig:fig3}
\end{figure*}

Figure~\ref{fig:fig3} shows a quantitative comparison of line-strengths of CO indices between data and different SSPs. For each index, the measurements on the stacked spectra of the XSGs are plotted with red circle and triangle (corresponding to apertures of 1.2~kpc and $<$R$_{\rm e}$>=3.9~kpc, respectively), while the green circle is for NGC~1277. The error bar on the red circle indicates the standard deviation of the measurements on the spectra of individual galaxies (note that they are extracted within the seeing-limited inner aperture of $\pm$0.675\arcsec, see \citetalias{labarbera2019}). E-MILES model line-strengths of the COs as a function of age, for solar metallicity and MW-like IMF SSPs, are shown with a solid pink line. Tension between the observed and model CO indices can be seen clearly. Since stars in the AGB phase account for nearly half of the K-band luminosity for stellar populations with ages of 0.1-2~Gyr \citep{maraston1998}, presence of intermediate-age stellar populations has been suggested as a source for the deep CO bandheads in the K-band spectrum of ETGs \citep{mobasher1996,james1999,mobasher2000}. However, \citetalias{eftekhari2022} showed that stellar population models with an enhanced contribution of AGB stars, constructed by adding a fraction of intermediate-age stellar populations on top of a dominant old population (blue arrow), do not match the observed CO indices. Indeed, the effect is so small and it is about the same order as the change due to metallicty variations (dotted pink line shows the predictions of E-MILES SSPs with a metallicity higher than solar ([M/H]=0.26)). Only for the CO1.64 and CO2.30 indices the change is slightly larger than the effect of metallicity. \citetalias{eftekhari2022} suggested an alternative scenario where massive ETGs might host rather old stellar populations, consistent with results from the optical range, but with enhanced [C/Fe] abundance. In Fig.~\ref{fig:fig3}, the violet arrow from \citetalias{conroy2018} models shows this effect. Although not enough to match the indices, however the impact of carbon-enhancement on CO indices is larger than the two other effects (AGB stars from intermediate-age populations and metallicity), especially on CO1.58, CO1.60, and CO1.64 indices, making the models' predictions significantly closer to observations (see \citetalias{eftekhari2022} for a detailed discussion on these effects). 
    
In Fig.~\ref{fig:fig3}, we also show, with a khaki arrow, the empirically corrected version of E-MILES SSP models constructed by relying on CO-strong cool giants (see \citetalias{eftekhari2022}) for a MW-like IMF.  This arrow approaches the observed line-strengths of the relic galaxy for CO1.56, CO1.58, and CO2.30, and matches the observations of CO1.60 and CO2.30 indices of the stacked spectra of XSGs. However, stellar population studies in the optical show that these  galaxies have a bottom-heavy IMF in the central regions ($\Gamma_{b}$$\sim$3; \citetalias{martin2015d}). Therefore, in the figure, we also include predictions of empirically-corrected E-MILES models with a bottom-heavy IMF. The orange arrow shows the effect of our empirical correction on bottom-heavy IMF models (solid and dotted purple lines for solar and metal-rich populations, respectively). Indeed, the figure shows that a bottom-heavy IMF implies shallower CO line-strengths, increasing the disagreement between models and observations. 
    
All panels in Fig.~\ref{fig:fig3} show that the mismatch between observations and models is not specific to the massive XSGs but it is also seen in the relic galaxy, NGC~1277. For the CO1.64 index, this mismatch is the same for both samples. On the other hand, NGC~1277 shows slightly stronger CO1.60 and CO1.66 line-strengths and shallower CO1.56, CO1.58 and CO2.30 line-strengths than the XSGs. The small difference between the CO line-strengths of NGC~1277 and XSGs, compared to the difference between observations and models, is not surprising as they do not share  exactly the same stellar population properties. For instance, while XSGs show a steep IMF gradient from the center to $\sim$4~kpc ($\Gamma_{b}$ from $\sim$3.0 to $\sim$1.3; \citetalias{labarbera2019}), the IMF of NGC~1277 only changes mildly from  the center to R$_{\rm e}$=1.2~kpc (from $\sim$3.0 to $\sim$2.5; \citetalias{martin2015d}). Moreover, the [Mg/Fe] is between 0.2-0.4~dex among XSGs (\citetalias{labarbera2019}) while it is around 0.4~dex for NGC~1277 (\citetalias{martin2015d}). Note also that the measurements of the CO indices on the stacked spectra of XSGs with different apertures are the same within the error bars, pointing to a negligible effect of stellar population gradients of XSGs in our analysis. The discrepancy between observations and model predictions, in both the XSGs and NGC~1277, rules out a scenario whereby  the CO absorptions are driven by  cool AGB stars from intermediate-age populations, as NGC~1277 is devoid of such a population. 
    
\section{Discussion}\label{sec:sec5}

The massive galaxies studied in this Letter show significantly stronger CO absorptions with respect to predictions of stellar population models. Early studies of ETGs in the NIR attributed the strong mismatch between observations of CO bandheads and models' predictions to the presence of intermediate-age stars in these galaxies (see Sec.~\ref{sec:sec1}), though this could arise some tension with respect to optical studies, finding evidence for red and dead populations in ETGs. Compact massive galaxies at high redshift are believed to be the remnants of primordial galaxies (z$\sim$3-5) that have formed the bulk of their stars in a short dissipative event with extremely high star formation rates and have undergone subsequent cessation of star formation, passively evolving up to z$\sim$2-3,  becoming the so-called ``red-nuggets". Normally, the red nuggets would experience mergers and gas inflows across cosmic time, increasing their size by a factor of $\sim$4 \citep{trujillo2006,vandokkum2008} and ending up as local giant ETGs. This might be the case for our sample of XSGs. They are classified as brightest cluster galaxies (BCGs) (except one of them) which could have experienced some (small level of) recent star formation as a result of accreting stars and gas from satellite galaxies orbiting around them and falling in. As a result, the XSGs could contain some fraction of intermediate-age stellar populations, especially in their central regions, consistent with the analysis of NUV spectral indices \citep{salvador2021}. However, no direct proof of the connection between the strengths of their CO absorptions and the presence of these younger populations has been provided so far. 

On the other hand, due to the stochastic nature of mergers, only a small fraction of red-nuggets can remain untouched until z$\sim$0, without experiencing a significant merger or interaction. Thus the so-called ``relic" galaxies are perfect laboratories to study the ``in-situ" stellar population of massive old ETGs. Since no rejuvenation of their stellar content has occurred since their early collapse phase, they are not supposed to host intermediate-age stellar populations, in contrast to BCGs. NGC~1277 has been identified as a ``true" relic system after its stellar populations have been investigated in detail by \citet{trujillo2014}. It is made up of a uniformly old stellar population ($>$10~Gyr) without a recent (i.e. $<$10~Gyr) star formation event \citep{ferre2015}. Note that \citet{beasley2018} constrained the accreted mass in NGC~1277 to be $\lesssim$12\% of its present day stellar mass while normal ETGs of the same stellar mass are expected to have accreted 50-80\% of their stellar mass.  Therefore, comparing the CO absorptions of this galaxy with the sample of BCGs provides us with a unique opportunity to test the long-believed explanation of the strong CO line-strengths origin in massive ETGs, i.e. due to the presence of intermediate-age populations. 

We measured \textit{H}- and \textit{K}-band CO line-strength indices, defined in \citetalias{eftekhari2021}, in both samples of massive ETGs and the relic galaxy and compared these measurements with predictions of stellar population models. Although our samples of BCGs and the relic galaxy have experienced different SFHs, leading to different amount of intermediate-age stellar populations (almost zero in the latter), the CO indices in both systems are significantly stronger than the predictions of the models. This points to the fact that intermediate-age stellar populations in massive ETGs are not responsible for the strong absorption of CO indices in these galaxies \citepalias{eftekhari2022}.  

Our result is also confirmed when using stellar population models to analyze the effect of other stellar evolutionary effects (C-enhancement and overweighting CO-strong giant stars; violet and orange/khaki arrows in Fig.~\ref{fig:fig3}, respectively) and the impact of SFH (an enhancement in the contribution of intermediate-age stellar populations; blue arrow in Fig.~\ref{fig:fig3}) on CO line-strengths. The former has a significant larger impact on CO strengths than the latter. In particular, to match the observations with models including a young component, the fraction of intermediate-age/young stellar populations should be so large but this is totally in disagreement with results inferred from optical studies of ETGs, and would prevent us from fitting the Balmer line-strengths.  Hence, our results rule out the scenario in which a significant contribution of intermediate-age stellar populations (i.e. an extended SFH) is needed to explain the strong CO absorption in the spectra of giant ETGs, reconciling the possible tension between the galaxy evolution picture inferred from optical and NIR spectral ranges. 

We remind the reader that there are some caveats into the comparison between observations of massive ETGs and stellar population models in the NIR because of SPS modelling uncertainties. First, empirical stars used to construct the models are all from the solar neighbourhood, having an unavoidable bias towards solar metallicity. While this makes the model predictions less reliable at high metallicity , as discussed in \citepalias{eftekhari2022},  the effect is not able to explain the CO mismatch. Second, current C-enhanced SPS models are based on theoretical stars that struggle to reproduce atomic and molecular bands for cool stars (T$\rm_{eff}$ < 4000~K). Thus, the effect of C-enhancement on models remains uncertain.

It is noteworthy to mention that our analysis in this Letter has been based on the comparison of a stacked spectrum of massive ETGs and that for an individual relic galaxy. In the future, we plan to enlarge the sample of relics at z$\sim$0,  in order to perform a comprehensive analysis of the differences between the two samples, and assess the scatter of CO line-strengths  in the population of relic  and non-relic galaxies. 

\section*{Acknowledgements}

We are thankful to the reviewer, Bahram Mobasher, for his valuable comments. We thank Ignacio Ferreras for useful comments and Jesus Falc{\'o}n-Barroso, Ignacio Mart{\'{\i}}n-Navarro, Marc Balcells, and Reynier Peletier for conversations that lead to the observation of NGC~1277. The authors acknowledge support from grant PID2019-107427GB-C32 from the Spanish Ministry of Science, Innovation and Universities (MCIU). This work has also been supported through the IAC project TRACES which is partially supported through the state budget and the regional budget of the Consejer\'\i a de Econom\'\i a, Industria, Comercio y Conocimiento of the Canary Islands Autonomous Community. FLB aknowledges support by the grant PRIN-INAF-2019 1.05.01.85.11. 

\section*{Data Availability}

Raw spectroscopic data for NGC~1277 (programme GTC3-17B; PI: MB) are available from the GTC Public Archive (\url{http://gtc.sdc.cab.inta-csic.es/gtc/jsp/searchform.jsp}). The XSGs  \citep{labarbera2019} were observed with ESO Telescope at the Paranal Observatory under programmes ID 092.B-0378, 094.B-0747, 097.B-0229 (PI: FLB) and data can be downloaded from the ESO archive (\url{http://archive.eso.org}). The E-MILES SSP models are publicly available at the MILES website (\url{http://miles.iac.es}). The AGB-enhanced and empirically corrected E-MILES models are available upon request to EE. The \citet{conroy2018} models are available for download at \url{https://scholar.harvard.edu/cconroy/sps-models}.  



\bibliographystyle{mnras}
\bibliography{references} 







\bsp	
\label{lastpage}
\end{document}